\documentclass{article}
\usepackage{natbib}
\usepackage{gcdv}
\usepackage{psfig}
\begin{document}
\small
\shorttitle{Galaxy Genesis}
\shortauthor{J.\ Bland-Hawthorn and K.C.\ Freeman}

\title
{\large \bf
Galaxy Genesis -- unravelling the epoch of dissipation in the
early disk
}

\author{\small 
 Joss Bland-Hawthorn$^1$ and Ken Freeman$^2$
} 

\date{}
\twocolumn[
\maketitle
\vspace{-20pt}
\small
{\center
$^1$Anglo-Australian Observatory, 167 Vimiera Road, Eastwood, NSW 2122, 
Australia\\
jbh@aao.gov.au\\[3mm]
$^2$Mount Stromlo Observatory, Australia National University, Weston Creek, ACT
 2611, Australia\\
kcf@mso.anu.edu.au\\[3mm]
}

\begin{center}
{\bfseries Abstract}
\end{center}
\begin{quotation}
\begin{small}
\vspace{-5pt}

So how did the Galactic disk form and can the sequence of events ever
be unravelled from the vast stellar inventory?  This will require that
some of the residual inhomogeneities from prehistory escaped the
dissipative process at an early stage. Fossil hunting  to date has
concentrated mostly on the stellar halo, but a key source of
information will be the thick disk.  This is believed to be a 'snap
frozen' relic which formed during or shortly after the last major epoch
of dissipation, or it may have formed from infalling systems early in
the life of the Galaxy.  As part of the KAOS Galaxy Genesis project, we
explore the early history of the halo and the thick disk by
looking for discrete substructures, either due to infall or in situ
star formation, through chemical tagging. This will require high signal
to noise, echelle spectroscopy of up to a million stars throughout the
disk. Our program has a short-term and a long-term goal.

The short-term goal is to quantify the size and structure of the
multi-dimensional chemical abundance space (${\cal C}$-space) for all major
components of the Galaxy. We seek to establish how many axes in ${\cal C}$-space
are decoupled and have large intrinsic dispersions. A critical test of
chemical tagging in the short term is that stellar streams in the halo,
identified from detailed phase space information, are highly localized
in ${\cal C}$-space, or are confined to chemical tracks. These trajectories
presuppose that stars form in a closed box through progressive
enrichments of the gas, leading to stars dispersed along a narrow track
in a complex chemical space. The long-term goal is to identify unique
chemical signatures in the thick disk, originating from different
formation sites, for star clusters which have long since dispersed.
This will require precise chemical abundances for heavy elements such
that a star can be localized to a discrete point in ${\cal C}$-space.  If the
star clusters originally formed outside the Galaxy in a bound infalling
system, the stellar abundances may fall along a chemical track, rather
than a discrete point in ${\cal C}$-space.
\end{small} \end{quotation} ]

\bigskip

\section{Introduction}

At the heart of a successful theory of galaxy formation must be a
detailed physical understanding of the dissipational processes which
form spiral galaxies.  The disk is the defining stellar component of
disk galaxies, and understanding its formation is in our view the most
important goal of galaxy formation theory.  Although much of the
information about the pre-disk state of the baryons has been lost in
the dissipative process, some tracers are likely to remain.

What do we mean by the reconstruction of early galactic history? We
seek a detailed physical understanding of the sequence of events which
led to the Milky Way. Ideally, we would want to associate components of
the Galaxy to elements of the protocloud -- the baryon reservoir which
fueled the stars in the Galaxy.

For many halo stars, and some outer bulge stars, this may be possible
with phase space information. But for much of the bulge and the disk,
secular processes cause the populations to become relaxed (i.e.  the
integrals of motion are partially randomized).  In order to have any
chance of unravelling disk formation, we must explore chemical
signatures in the stellar spectrum.

The exponential thin disk, with a vertical scale height of about 300
pc, is the most conspicuous component in edge-on disk galaxies.  The
thin disk is believed to be the end product of the quiescent
dissipation of most of the baryons and contains almost all of the
baryonic angular momentum. Many disk galaxies show a second fainter
disk component with a longer scale height (typically about 1 kpc);
this is known as the thick disk \citep{sd00,db02}.
 The Milky Way has a significant thick
disk \citep{gr83}:  its surface brightness is about 25\% of
the thin disk's surface brightness, and its stars are significantly
more metal poor than the stars of the thin disk.

The galactic thick disk is currently believed to arise from heating of
the early stellar disk by one or more accretion events although other
possible origins have been discussed \citep{gwk89,gwj95}.
It is composed
of only old stars, with ages greater than 10 Gyr, equivalent to forming
at a redshift of $z\geq1$ \citep{rw00}. Furthermore,  the thick disk is
chemically distinct from the thin disk at low metallicity 
\citep{kf98,fbl03}, in the sense that
[$\alpha$/Fe] is enhanced relative to the thin disk, although solar
values are seen at higher metallicity. The thick disk may be one of the
most significant components for studying signatures of galaxy formation
because it presents a 'snap frozen' relic of the state of the (heated)
early disk.

In the discussion which follows, we make reference to the KAOS project
(see http://www.noao.edu/kaos) which proposes a highly multiplexed
wide-field multi-object spectrograph for the Gemini Observatory. Full
specifications for the proposed instrument can be found at the above
web site.

\section{Unravelling a dissipative process}

In order to follow the sequence of events involved in dissipation, we
propose that the critical components which need to be re-assembled are
the individual star clusters which formed at each stage. Since most
stars are born in dense clusters, the formation and evolution of
galaxies today must involve millions of discrete cluster events
throughout their history. We would like to establish the evolving mass
function of star clusters, their chemical composition, formation and
survival rate as a function of cosmic time. Galaxy-wide enrichment from
the fall-out of nuclear winds or mergers would be evident in the fossil
record of reconstructed star clusters, assuming these provide an
unbiased sampling of cosmic time regardless of the star formation
history.

But how are we to reconstruct star clusters which have long since
dispersed? It will be necessary to tag individual stars to their parent
cloud through unique chemical signatures shared by these stars,
assuming these exist.

We now discuss our basic strategy for 'chemical tagging' 
\citep[][hereafter FBH]{fb02}. High resolution spectroscopy at
high signal-to-noise ratio of many stellar types reveals an
extraordinarily complex pattern of spectral lines.  The spectral lines
carry key information on element abundances that make up the stellar
atmosphere. Many of these elements cannot arise through normal stellar
evolution, and therefore must reflect conditions in the progenitor
cloud at the time of its formation.

\subsection{Primary requirements of chemical tagging.}

Our long-term goal is to chemically tag stars into coeval groups, i.e.
to identify individual members of star clusters which have long since
dispersed. For unique chemical signatures to exist, there are several
key requirements (FBH):

1. Most stars must be born in dense star clusters.

2. Most dense star clusters must be chemically uniform in key elements.

3. Key chemical elements must reflect the cloud composition of the
   progenitor cloud.

4. Key chemical elements must not be rigidly coupled (i.e. vary in lock
step), and there must be sufficient abundance dispersion in key
elements to allow for unique groups (reflecting unique sites of
formation) to be readily identified.

5. There must be a contiguous spectral window which contains the
necessary information on key elements for chemical tagging.

We discuss each of these requirements, or conditions, in turn.
Conditions 1, 2, 3 and 5 appear to be supported by observation.
Condition 4 is the most uncertain largely because stellar abundance
surveys to date target either too few stars or too few chemical
elements.

{\bf Condition 1.}    Most stars are born within rich clusters of many
hundreds to many thousands of stars \citep{cbh00,jc00}.
This essential fact is supported by many studies from optical,
infrared, millimeter and radio surveys.

{\bf Condition 2.}    The widely held view that open clusters are
chemically uniform can be traced to classic work on Hyades by 
\citet{cww65}. But, until recently, very little was known about detailed
heavy element abundance work on open clusters with rigorous membership
established by  reliable astrometry or accurate radial velocities ($<$
0.5 km/s), so as to minimize 'pollution' from stars not associated with
the cluster \citep[e.g.][]{aq02}.

There have been several recent studies of open clusters which have
established that open clusters are indeed highly uniform. In a detailed
analysis of 55 F$-$K dwarfs in Hyades, 
\citet{psc03} find that
the abundance variations in Si, Ti, Na, Mg, Ca and Zn with respect to
Fe, and in [Fe/H], are within the measurement errors ($<$0.04 dex).

There is remarkably little theoretical work which addresses the
chemical uniformity of star clusters. In theoretical work on star
formation, chemical uniformity is almost always a prior assumption. Are
we to expect chemical uniformity among low mass stars in an open
cluster? To date, this question has not been adequately answered.

\citet{mt02} propose that high-mass stars form in the cores of
strongly self-gravitating and turbulent gas clouds. Two possible routes
to chemical uniformity is that all stars form at the same instance from
a chemically uniform cloud, or that the low mass stars form outside of
the core shortly after the supernovae have uniformly enriched the
cloud. The precise sequence of events which give rise to open clusters
is a topic of great interest and heated debate in contemporary
astrophysics \citep[e.g.][]{sph00}.

{\bf Condition 3.}    It is generally believed that r-process elements
(e.g.  Sm, Eu, Gd, Tb, Dy, Ho) cannot be formed during quiescent
stellar evolution.  While some doubts remain, the most likely site for
the r-process appears to be Type II supernova (SN II), 
as originally suggested by 
\citet{bbfh57} \citep[see also][]{gw97}. Therefore, r-process
elements measured from stellar atmospheres reflect conditions in the
progenitor cloud. The same is believed to be true for most of the
$\alpha$ elements since these are produced in the hydrostatic burning
phase of the pre-supernova star.

In contrast, the s-process elements (e.g.  Sr, Zr, Ba, Ce, La, Pb) are
thought to arise from the He-burning phase of intermediate to low mass
(AGB) stars (M$<$10 M$_\odot$), although at the lowest metallicities,
trace amounts are likely to arise from high mass stars \citep{bpa00,rhhw02}.

{\bf Condition 4.}   During the past four decades, evidence has
gradually accumulated for a large dispersion in metal abundances [X/Fe]
(particularly n-capture elements) in low metallicity stars relative to
solar abundances 
\citep{wgpha63,bp65,ss78,jt81,lb85,dc88,gspc88,mps95,nrb96,bpa00}.
  Elements like Sr, Ba and Eu show a 300-fold dispersion 
\citep[see Fig.\ 21 of][]{gw97},
although [$\alpha$/Fe] dispersions are typically an order of
magnitude smaller.

These observations have been used to constrain detailed supernova
models, which in turn show how different yields arise as a function of
progenitor mass, progenitor metallicity, mass cut (what gets ejected
compared to what falls back towards the compact central object), and
detonation details.  These models help to explain the smaller
dispersions in $\alpha$ elements: the $\alpha$ yields are not dependent
on the mass cut or details of the fallback/explosion mechanism, which
leads to a smaller dispersion at low metallicity.

Is there evidence that some elements are decoupled? The element
abundances [X/Fe] show three main peaks at Z$\sim$26, Z$\sim$52, and
Z$\sim$78 \citep{scb01,slc01}. There have been suggestions that the
r-process gives rise to random abundance patterns 
\citep[e.g.][]{ga96}, but this is not supported by new observations of a few
metal poor stars.  Heavy r-process elements around the second peak
compared to the Sun appear to show a universal sawtooth pattern, albeit
with some internal variance, in the range 56$<$Z$<$75 
\citep{sci00,chb01,hcb02}.
 But \citet{omk03}
stress that this does not imply a universality of abundances outside
this range, and in fact this is not observed. \citet{hcb02} find that the
third peak and actinide elements (Z$\geq$90) are decoupled from
elements in the second peak. There may be a substantial number of
suitable elements (10-20) which could define a sufficiently large
parameter space.

 The very large scatter of [n-capture/Fe]
in ultra-metal poor stars seen in Figure 12 of \citet{gw97} 
means that n-capture element abundances in ultra-metal poor stars 
are products of one or very few
prior nucleosynthesis events that occurred in the very early, poorly
mixed galactic halo, a theme that has been developed by many authors
\citep[e.g.][]{gspc88,as95,mps95,st98,asgt00,tsy00}.

\begin{figure} 
\psfig{file=blank.ps,width=7cm}
\caption{
An illustration of how successive chemical enrichments by supernovae can
 lead to low mass stars with a wide scatter in chemical abundances
 which converge to a universal  value.  The model assumes a constant
 star formation rate in a closed box. The stars are formed according to
 a Salpeter mass function and the yields are taken from models by
 \citet{ts98}. Only 1\% of the stars produced in the simulation are 
shown after the initial burst; we show all the stars at t=0. }
\label{eufe-fig}
\end{figure}

The simulation in Figure \ref{eufe-fig} 
is illustrative of the expected abundance
variation in successive generations of low mass stars in a closed box
from gas enriched by successive generations of supernovae. The upper
envelope is determined by enrichment from low mass supernovae, the
lower envelope from high mass supernovae. From left to right, the black
squares indicate the number of enrichment events, i.e. 1, 10, 100, 1000
events. Just ten enrichments from high mass supernovae are sufficient
to enrich a cloud to [Fe/H]$=-2$. Note the rapid convergence in this
simple model above [Fe/H]$=-1$.

In the simulation, we consider only core collapse supernovae;
subsequent enrichment by Type Ia supernovae causes the converging
stream to dip down above [Fe/H]$=-1$. It is not clear at what [Fe/H]
the r-process elements become swamped by the ubiquitous Fe-group and
s-process elements. \citet{tgg99} suggest that the s-process
does not become significant until [Fe/H]$\sim-1$ because of the need
for pre-existing seed nuclei \citep{ss78,tbcs81},
although \citet{pt97} argue for some s-process
production at [Fe/H]$\sim-2.5$. In any event, Figure \ref{asspec1-fig} 
makes clear
that, at sufficiently high resolution, both the Sun and Arcturus reveal
a rich network of heavy elements, many of which arise from the
r-process.

An important point to appreciate is that the substantially smaller
scatter in [$\alpha$/Fe] \citep[see][]{cgcbc02}, compared to heavy
element scatter, presumably argues that the scatter is unlikely to
arise from chemical differentiation, i.e. where metal yields from a
given supernova are distributed in different amounts, compared to Fe,
to the surrounding ISM. In other words, the mixing is sufficiently
uniform such that any given cell receives the same proportion of Eu/Fe
from the supernova. This puts limits on the degree to which asymmetric
supernovae \citep{mnn02} must differentially enrich the
surrounding gas.

\begin{figure*}[t]
\begin{tabular}{cc}
\psfig{file=blank.ps,width=7cm} &
\psfig{file=blank.ps,width=7cm} 
\end{tabular}
\caption{
 A demonstration of where most of the chemical information lies in a
 highly resolved spectrum of Arcturus (left) and the Sun (right).  The
black lines in the top plots give the total ion count in a spectral
 window $\Delta \lambda = \lambda/35$.  The magenta lines show what
fraction of this count is made up of elements with two ionization
states. The lower plots separate these into element types. Note the
rapid rise of heavy elements to the blue in both stars; note also that
the solar spectral information extends to bluer wavelengths.  }
\label{asspec1-fig}
\end{figure*}

{\bf Condition 5.}   Any practical experiment involving chemical
signatures in millions of stars will require a wide-field, multi-fibre
spectrograph operating at echelle resolutions. Since rayleigh
scattering renders optical fibres almost useless much below 4000 ${\rm
\AA}$, it is important to establish that there exist contiguous
spectral windows which contain abundance information on a large number
of chemical elements.  Since it will be necessary to maximize the number
of object fibres at the detector, this information will need to exist
within a single narrow band ($\sim\lambda$/35) echelle order. The
optimal parameters for a 'galaxy genesis' machine are determined below.

\begin{figure} 
\psfig{file=blank.ps,width=8cm,angle=-90}
\caption{
 17 ${\rm \AA}$ spectral window (4047-4064${\rm \AA}$) of the Sun
 obtained with the Fourier Transform Spectrograph at the National Solar
 Observatory revealing detailed chemistry on Fe, Cr, Ti, V, Co, Mg, Mn,
 Nd, Cu, Ce, Sc, Gd, Zr, Dy. The lower values are wavenumbers
 (cm$^{-1}$). The data are taken from http://www.nso.edu/.
 } 
\label{sspec-fig}
\end{figure}

Do there exist narrow band spectral windows, with chemical information
on dozens of elements, which could be utilized by a
fibre-fed echelle spectrograph? Until recently, this has not been
possible to answer objectively.  Figure \ref{sspec-fig} 
 shows such a window that
was revealed by this study.  An optimal window needs information on
more than one ionization state for as many elements as possible in
order to accurately determine the stellar surface gravity. Various
ionic transitions in combination must also provide accurate information
on the stars luminosity and surface temperature. Some regions of the
spectrum are dominated by telluric features; in cool stars, molecular
bands complicate the spectrum.

We concentrate our analysis on digitized atlases of the Sun
([Fe/H]$=$0.0) and for the luminous giant Arcturus ([Fe/H]$=-0.56$)
\citep{mrs03}. Our solar line list derives from data supplied
by P. Hall (NOAO) transcribed from the \citet{mmh66}
solar spectrum (2935-8770\AA). The atlas includes information on
25,000 absorption lines covering 63 elements, for which 28 are detected
in two ionization states. The Arcturus line list comes from the
digitized atlas (3570-7405\AA) of \citet{whl98}.
The line list includes 7700 atomic transition for 42 elements, for
which 18 are detected in two ionization states. Both lists are
dominated by Fe I (45\%) and light $\alpha$ elements (30\%).

Figure \ref{asspec1-fig} shows the number of unique ions detectable as a function of
wavelength for the Sun and for Arcturus. At each wavelength, the
accessible window is $\Delta\lambda$ where
$\lambda$/$\Delta\lambda=$35, such that at $\lambda=3500\ {\rm \AA}$,
the unique ion count is made over a 100 ${\rm \AA}$. The linear
dependence of the window on wavelength is expected for spectral
coverage in a single echelle order limited by the detector size. Note
the rising count towards with decreasing wavelength down to 4000 ${\rm
\AA}$, both in terms of the number of unique ions, and elements with
two ionization states.

In the lower panels of Figure \ref{asspec1-fig}, we demonstrate that this rise is largely
due to a rising fraction of neutron capture elements; the rise in iron
peak elements is not as dramatic. It is noteworthy that we obtain
similar results for a solar metallicity main sequence star and a metal
weak giant. Of particular importance to the KAOS project, note that
there is an optimal window close to 4100 ${\rm \AA}$ accessible to
optical fibres.

\begin{figure*}[t]
\begin{tabular}{cc}
\psfig{file=blank.ps,width=7cm} &
\psfig{file=blank.ps,width=7cm} 
\end{tabular}
\caption{
The effect of resolving power on the fraction of spectral lines which
can be resolved as a function of wavelength. In order to resolve 80 \%
of lines at 4000 ${\rm \AA}$, we need R = 40,000; this falls to R =
10,000 at 8000 ${\rm \AA}$. Once again, both stars reveal the same
general trends; note also that the solar spectral information extends
to bluer wavelengths.  } 
\label{asspec2-fig}
\end{figure*}

In Figure \ref{asspec2-fig}, 
we show the effect of resolving power on the fraction of
lines which can be resolved as a function of wavelength. The optimal
window at 4100 ${\rm \AA}$ requires a spectral resolving power of R =
40,000 in order to resolve 80 \% of the lines. As the resolving power
decreases, the loss of information on heavy elements is particularly
dramatic.

We conclude that the KAOS Galaxy Genesis project should target 4100
${\rm \AA}$ in a 100 ${\rm \AA}$ window at a resolving power of R =
40,000. The required sampling 0.05 ${\rm \AA}$ per pixel or 2000 pixels
in the dispersion direction, well matched to state of the art CCDs.

\subsection{Candidates for chemical tagging}

Chemical tagging is not possible for all stars.  In hot stars and young
star clusters, our ability to measure abundances is reduced by the
stellar rotation and lack of transitions for many ions in the optical.
In very cool stars, very little light emerges in the optical or near-IR
due to complex molecular and dust opacities. The ideal candidates are
evolved FGK stars that are intrinsically bright, FGK subgiants and
dwarfs. These populations probably account for 10\% of all stars.

Giants can be observed at R = 40,000 over the full Gaiasphere, i.e. the
20 kpc diameter sphere centred on the Sun which will be surveyed in
great detail by the GAIA satellite. Dwarf stars will only be observable
within 1-2 kpc. While this is only a small fraction of the available
volume, the local volume may include a representative sample
of all old disk stars, regardless of their point of origin.

For many stars, certain spectral regions are dominated by thermally
broadened profiles, particularly for light elements. In general,
heavier atoms show narrow profiles appropriate to their mass, although
odd numbered atomic isotopes are susceptible to hyperfine splitting
(due to the non-zero nuclear magnetic moment) which produces a
broadened asymmetric line profile.

Giants have deep, low density atmospheres that produce strong
low-ionization absorption lines compared to higher gravity atmospheres.
Even in the presence of significant line blending, with sufficient
signal, it should be possible to derive abundance information by
comparing the fine structure information with accurate stellar
synthesis models. Detailed abundances of large numbers of F and G
subgiants would be particularly useful, if it becomes possible to make
such studies, because direct relative ages can be derived for these
stars from their observed luminosities.

\section{Short-term goal -- size and structure in a multi-dimensional ${\cal C}$-space}

An intriguing prospect is that reconstructed star clusters can be
placed into an evolutionary sequence, i.e.  a family tree, based on
their chemical signatures. Let us suppose that a star cluster has
accurate chemical abundances determined for a large number n of
elements (including isotopes). This gives it a unique location in an
n-dimensional space compared to m other star clusters within that
space. We write the chemical abundance space as ${\cal C}$(Fe/H,
X$_1$/Fe, X$_2$/Fe, ...) where X$_1$, X$_2$ ... are the independent chemical elements that define
the space (i.e. elements whose abundances are not rigidly coupled to
other elements).

Our simple picture assumes that a cloud forms with a unique chemical
signature, or that shortly after the cloud collapses, one or two
massive SN II enrich the cloud with unique yields which add to the
existing chemical signature. The low-mass population forms with this
unique chemical signature. If the star-formation efficiency is high
($\geq$30\%), the star group stays bound although the remaining gas is
blown away. If the star-formation efficiency is low, the star cluster
disperses along with the gas. In a closed box model, the dispersed gas
reforms a cloud at a later stage.

In the closed box model, each successive generation of supernovae
produce stellar populations with progressive enrichments. These will
lie along a trajectory in ${\cal C}$-space \citep[see Fig.\ 8 of ][]{kg01}.
 The overall
distribution of the trajectories will be affected by fundamental
processes like the star formation efficiency, the star formation
timescale, the mixing efficiency, the mixing timescale, and the
satellite galaxy infall rate.

A critical test of chemical tagging is that stellar streams in the
halo, identified from detailed phase space information, are highly
localized in ${\cal C}$-space, or are confined to chemical tracks  
this is a key
short term goal. There may already be evidence for accreted halo stars
from their distinct chemical signatures (e.g. suppressed Mg) since
those with estimated orbital parameters are found to have large
apogalacticon distances \citep{jk97,ns97,cws97,cgcbc02}.
 This may suggest that these stars
originated in lower mass stellar systems with very different chemical
histories from that of the Galaxy.

Our team (De Silva, Bland-Hawthorn, Freeman, Bessell, Asplund) is
currently engaged in large surveys to establish the chemical
homogeneity of open clusters as this forms a central tenet of chemical
tagging. We propose to extend this survey to large star-forming
complexes which include several stellar associations within one or two
super-associations. This is the sample volume we need to tackle to
establish the existence of chemical tracks. Presently, it is difficult
to identify a proper sample until phase space information becomes
available from GAIA, although Hipparcos has provided detailed candidacy
for a few star forming complexes \citep{dhd01}.

As we approach solar levels of metallicity in [Fe/H], the vast number
of trajectories will tend to converge \citep[see Fig.\ 21 of][]{gw97}.
By [Fe/H]$\sim-$2.5, AGB stars will have substantially raised the s-process element
abundances; by [Fe/H]$\sim -$1, Type Ia supernovae will have raised the
Fe-group abundances.  Star clusters that appear to originate at the
same location in this ${\cal C}$-space may simply reflect a common formation
site, i.e.  the resolution limit we can expect to achieve in
configuration space.

Even with a well established family tree based on chemical trajectories
in the chemical ${\cal C}$-space, this information may not give a clear
indication of the original location within the protocloud or Galactic
component. This will come in the future from realistic baryon
dissipation models. Forward evolution of any proposed model must be
able to produce the observed chemical tree.

However, the ${\cal C}$-space will provide a vast amount of information on
chemical evolution history. It should be possible to detect the
evolution of the cluster mass function with cosmic time \citep{pk02},
the epoch of a starburst phase and/or associated mass ejection of
metals to the halo \citep{ar02}, and/or satellite infall 
\citep{mn98}. The chemical tracks could conceivably be punctuated by
discontinuities due to dramatic events like galactic infall or
large-scale winds \citep{cmbn99}.

As we go back in time to the formation of the disk, we approach the
chemical state laid down by population III stars.  The rarity of stars
below [Fe/H]$\sim -$5 suggests that the protocloud was initially
enriched by the first generation of stars \citep{asgt00} or
maybe that stars moving through the ISM today have a minimum threshold
metallicity due to Bondi-Hoyle accretion. However, the apparent absence
of any remnants of population III remains a puzzle: its stars may have
had a top-heavy initial mass function, or have dispersed into the
intra-group medium of the Local Group.  If one could unravel the
abundances of heavy elements at the time of disk formation, this would
greatly improve the precision of nucleo-cosmochronology. Important
information is beginning to emerge from echelle observations of damped
Ly$\alpha$ systems at high redshift \citep{pmb03}.

\subsection{The size of ${\cal C}$-space}

The size of n is unlikely to exceed about 60 for the foreseeable
future. \citet{scl03} present exquisite data for the metal-poor
star CS 22892-052, where abundance estimates are obtained for a total
of 54 elements, with tight upper limits on 3 other elements. 
\citet{hcb02} present exquisite data for the metal-poor star CS
31082-001, where abundance estimates are obtained for a total of 44
elements, almost half the entire periodic table 
\citep[see also][]{chb01,scb01,slc01}.
 The $\alpha$ and r-process
elements, and maybe a few canonical s-process elements at low [Fe/H],
provide information on the cloud abundances prior to star formation,
although combinations of these are likely to be coupled 
\citep{hw01,scb01,slc01}.  There are more than two dozen
r-process elements that have been clearly identified in stellar spectra
\citep{gw97,scl03}.

The size of m is likely to be exceedingly large for the thin disk where
most of the baryons reside. For a rough estimate, we take the age of
the disk to be 10 Gyr. If there is a unique SN II enrichment event
every 100 years, we expect of order 10$^8$ formation sites.  Typically, a
SN II event sweeps up a constant mass of 5$\times10^4$ \citep{rnb96,st98}.
 Simple chemical evolution models
indicate that this must be of the right order to explain the
metallicity dispersion at low [Fe/H] \citep{asgt00}. Roughly
speaking, there have been 10$^3$ generations of clouds since the disk
formed, with about 10$^5$ clouds in each star-forming generation, such
that  cloud formation and dispersal cycles on a 10$^7$ yr timescale
\citep{eepz00}.

Whereas the total number of star clusters over the lifetime of the thin
disk is very large, the size of m for the stellar halo 
\citep{hmo01}, and maybe the thick disk \citep{pk02}, is likely to be
significantly smaller.  Our primary interest is the oldest star
clusters.  Reconstructing star clusters within the thick disk is a
particularly interesting prospect since the disk is likely to have
formed within 1-1.5  Gyr of the main epoch of baryon dissipation
\citep{pncmw00}.

\subsection{How many unique chemical signatures make up the halo or the thick disk?}

There are different approaches to answering this question: (i) carry
out a cold dark matter simulation (with hydro) and trace stellar
evolution and feedback within each of the building blocks which make up
a galaxy; (ii) carry out a Press-Schechter simulation based on
statistical aggregation with prescriptive physical laws; (iii)
undertake an order of magnitude calculation.

Our basic view is that modern CDM simulations in (i) are not able to
carry enough chemical information in order to do this properly. In
its place, the numerical approach in (ii) is currently under
investigation by us: our initial
conclusion is that the expected number of signatures is very poorly
defined due to uncertainties in supernova yields, and the sequence of
events involved in star formation and recycling of gas. 
In place of detailed simulations, we provide an
order of magnitude estimate.

There are at least three possible scenarios for the thick disk, all of
which can be tested by chemical tagging. (1) The thick disk is thought
to be a snap-frozen relic of the early disk, heated vertically by the
infall of an intermediate mass satellite. This is a particularly
interesting prospect since chemical tagging would provide clues on the
formation of the first star clusters in the early disk. (2) \citet{pk02}
has suggested that the thick disk arose from a population of
super-star clusters ($\sim10^6$ M$_\odot$) which became unbound. (3)
\citet{bc00} have shown how the thick disk could
arise from the infall of a number of Local Group satellites; see
also \citet{anse03}.
Their models could explain why thick disk
stars are more metal rich than stars in dwarf galaxies or other
likely building blocks, because the minor merger would have triggered
star formation and self-enrichment.

For a thick disk made up of super-star clusters, we would need to
detect roughly 10$^4$ unique chemical signatures, requiring a survey of
10$^5$ stars in order to detect a unique chemical signature at the
3$\sigma$ level. For a thick disk made up of the heated early disk, we
would need to detect of order 10$^7$ unique signatures, requiring a
survey of 10$^8$ stars. For a thick disk comprising roughly ten Local
Group satellites, the number of unique chemical signatures will depend
on the details of the star formation history within these satellites
prior to infall. But very roughly, we would need to detect 10$^6$
unique signatures, requiring a survey of 10$^7$ stars.

In the next section, in a more detailed statistical analysis, we show
that a survey of 10$^6$ stars may be sufficient to identify the
chemical signatures of the early thick disk and the stellar halo.

\subsection{Searching for progenitor formation sites -- how many stars do we need?}

Our simulations show that we need to identify roughly 
10\% of the original formation sites in order to properly sample cosmic
time for a wide range of possible star formation histories (stochastic,
exponential). In the table below, we show the size of survey needed to
detect a certain fraction of formation sites where the detection
requires 1, 2, 3, 5, 10 or 30 stars from each site (Table \ref{tab1}).

\begin{table*}
\caption{
The number of stars needed to detect progenitor sites
(normalized to the number of sites). The rows indicate the fraction of
sites which are detected in 1, 2, 3, 5, 10, 30 stars. For example, if
there are 1000 possible sites, and we want to detect 5 stars in 10\% 
of these, we need to randomly sample about 2400 stars from the complete 
sample.
}
\label{tab1}
\begin{center}
\begin{tabular}{cccccccccc}
\hline
 & 10\% & 20\% & 30\% & 40\% & 50\% & 60\% & 70\% & 80\% & 90\%\\
\hline
 1 & 0.11 & 0.22 & 0.36 & 0.50 & 0.68 & 0.90 & 1.19 & 1.60 & 2.32\\
 2 & 0.53 & 0.81 & 1.08 & 1.36 & 1.67 & 1.99 & 2.41 & 2.97 & 3.87\\
 3 & 1.10 & 1.53 & 1.92 & 2.27 & 2.67 & 3.10 & 3.62 & 4.26 & 5.29\\
 5 & 2.43 & 3.09 & 3.61 & 4.13 & 4.65 & 5.21 & 5.88 & 6.73 & 8.08\\
 10 & 6.22 & 7.30 & 8.16 & 8.90 & 9.65 & 10.46 & 11.41 & 12.60 & 14.21\\
 30 & 23.20 & 25.30 & 26.92 & 28.33 & 29.64 & 31.05 & 32.69 & 34.56 &
 37.21\\ \hline
\end{tabular}
\end{center}
\end{table*}

{\bf Thick disk.} If we adopt \citet{pk02} model for the thick disk,
we would need to survey 60,000 thick disk stars in order to establish
3$\sigma$ chemical signatures for 10\% of the original formation sites.
As we show below, this requires 500$-$1000 KAOS pointings in order to
survey such a large number of thick disk stars in a limited magnitude
range.

{\bf Stellar halo.} The baryon mass of the stellar halo is in the 
range 10$^8$ to 10$^9$ M$_\odot$ depending on the details of how 
we distinguish between halo and bulge stars, and whether or not we
include recently identified halo objects (e.g. Sgr dwarf). For 
illustration, we take the lower bound here since we can scale 
trivially to the higher bound.

Early universe simulations suggest that the first star clusters (e.g.
protoglobulars, first dwarfs) must contain roughly a Jeans mass, or
about 10$^5$ M$_\odot$ in baryons \citep{abn02}.  Since a single
supernova can enrich a cloud of 10$^5$ M$_\odot$, let us suppose that
the Galactic halo arises from 1000 progenitor sites.  Therefore, we
would need to survey 500 halo stars to be sure of seeing the chemical
signatures of more than one star from 10 \% of the original formation
sites. With sufficiently accurate abundance information, a survey of
6000 halo stars would give a discrete chemical signatures (3$\sigma$)
for 10 \% of the original sites.

If the original halo formation sites saw several generations of star
formation before infall, we should expect to see chemical tracks in
${\cal C}$-space. If we need 30 stars from a single formation site to
establish a chemical track, we need to survey roughly 25,000 halo stars
in order to establish tracks for 10\% of the progenitor sites. This
amounts to 500-1000 KAOS pointings which can be observed simultaneously
with the thick disk pointings (see below).

If chemical tracks exist, it is fundamentally important to identify
them because the starting point identifies a unique set of chemical
signatures from a first-generation supernova. As mentioned above, there
are a host of fundamental uncertainties about the enrichment details
from supernovae, which severely limits our ability to perform realistic
CDM+hydro simulations. But there is likely to be substantial progress
in this arena \citep[e.g.][]{gfrkl03,bkgf04}.

Note that the chemical tagging of the halo will be assisted by phase
space information supplied by the KAOS survey and the GAIA space
mission. It is presently unclear whether  phase space information for
the thick disk will provide supporting evidence for discrete groups
identified by their unique chemical signatures.

In summary, a survey of 25,000 halo stars and 60,000 thick disk stars
may be sufficient to detect a useful number of progenitor formation
sites; these observations will share 500-1000 KAOS pointings (see
below).

\subsection{How many chemical signatures are we ever likely to detect? }

The data to answer this question do not exist: stellar surveys to date
target either too few stars or too few chemical elements. The answer
depends in part on the strategy: we could attempt accurate measurements
($<$0.05 dex) on a few decoupled lines, or try to distinguish, say, a
high and a low state in many elements.  An error of 0.02 dex is
equivalent to 1 ${\rm m\AA}$ measurement accuracy in equivalent width.

Most physical processes are intrinsically noisy, so it would be
surprising if there did not exist at least a small amount of decoupling
in elements which appear to correlate within the measurement errors
over a wide dynamic range in metallicity. The rapid statistical
convergence with increasing [Fe/H] seen in Figure 21 of \citet{gw97}
is assumed to arise from
the increasing homogeneity of the ISM with cosmic time. But some
intrinsic scatter could be sustained by gaseous infall 
\citep[however see][]{pt97}.

In Figure \ref{asspec1-fig}, 
there are more than 30 unique elements observable in our
optimal window. If we could distinguish a high and a low state for each
of these, we could in principle distinguish a billion unique chemical
signatures. Either strategy would require an accurate differential
abundance analysis \citep{eag93,pncmw00,psc03}.

But we stress that it may not be necessary to measure as many as 30
elements if some can be found which are highly decoupled and exhibit
large relative dispersions from star to star. \citet{bpa00}
demonstrate one such element pair, i.e. [Ba/Fe] vs [Sr/Fe], confirmed
by \citet{cgcbc02} who also suggest [Mn/Fe] vs [Cr/Fe] may have
appreciable dispersion. It is part of our short-term goal to establish
which axes in ${\cal C}$-space have the largest dispersions, and which
axes are likely to be most decoupled.

\section{Long-term goal -- reconstructing ancient star groups from unique chemical signatures}

The abundance dispersion in $\alpha$ and heavy elements provides a
route forward for tagging groups of stars to common sites of
formation.  With sufficiently detailed spectral line information, it is
feasible that the 'chemical tagging' will allow temporal sequencing of
a large fraction of stars in a manner analogous to building a family
tree through DNA sequencing.

Consider the (extraordinary) possibility that we could put many coeval
star groups back together over the entire age of the Galaxy. This would
provide an accurate age for the star groups either through the
color-magnitude diagram, or through association with those stars within
each group that have [n-capture/Fe]
 $\gg$ 0, and can therefore be radioactively dated.  This would provide
 key information on the chemical evolution history for each of the main
 components of the Galaxy.

There is no known age-metallicity relation that operates over a useful
dynamic range in age and/or metallicity. (This effect is only seen in a
small subset of hot metal-rich stars).  Such a relation would require
the metals to be well mixed over large volumes of the ISM.  For the
foreseeable future, it seems that only a small fraction of stars can be
dated directly (FBH).

Ideally, we would like to tag a large sample of representative stars
with a precise time and a precise site of formation. Can we identify
the formation site? The kinematic signatures will identify which
component of the Galaxy the reconstructed star group belongs, but not
specifically where in the Galactic component (e.g.  radius) the star
group came into existence. For stars in the thin disk and bulge, the
stellar kinematics will have been much affected by the bar and spiral
waves; it will no longer be possible to estimate their birthplace from
their kinematics.  Our expectation is that the derived family tree will
severely restrict the possible scenarios involved in the dissipation
process. In this respect, a sufficiently detailed model may be able to
locate each star group within the simulated time sequence.

Our ability to detect structure in ${\cal C}$-space depends on how
precisely we can measure abundance differences between stars. It may be
possible to construct a large database of differential abundances from
echelle spectra, with a precision of 0.05 dex or better; differential
abundances are preferred here to reduce the effects of systematic
error.

\begin{table*}
\caption{
Distance limit in log(parsecs) for different stars as a function of apparent V magnitude: metal poor giants (MPG), metal rich giants (MRG), clump giants (CG), blue horizontal branch halo (BHB), and main sequence dwarfs. The second column is the absolute V magnitude of the star. Brackets help to delineate the transition between 1 - 10 - 100 kpc. Note that the Solar Circle provides an extra 8 kpc in radial extent such that surveys which reach the Galactic Center also reach the outer disk.
}
\label{tab2}
\begin{center}
\begin{tabular}{cccccccccc}
\hline
 &  V  &       13.  &  14. &   15. &   16. &   17. &   18. &   19.  &  20.\\
\hline     
MPG & -2.0  & (4.0) &  4.2  &  4.4  &  4.6  &  4.8  & (5.0) &  5.2  &  5.4 \\
    & -1.5  &  3.9  &  4.1  &  4.3  &  4.5  &  4.7  &  4.9  &  5.1  &  5.3 \\
MRG & -1.0  &  3.8  & (4.0) &  4.2  &  4.4  &  4.6  &  4.8  & (5.0) &  5.2 \\
    & -0.5  &  3.7  &  3.9  &  4.1  &  4.3  &  4.5  &  4.7  &  4.9  &  5.1 \\
    &  0.0  &  3.6  &  3.8  & (4.0) &  4.2  &  4.4  &  4.6  &  4.8  &
 (5.0) \\
CG/BHB &  0.5  &  3.5 &  3.7 &   3.9  & 4.1 &  4.3 &  4.5 &  4.7 &  4.9 \\
       &  1.0  &    3.4 &  3.6 &  3.8 & (4.0) & 4.2 &  4.4 &  4.6 &  4.8 \\
       & 1.5  &    3.3 & 3.5  & 3.7 &  3.9 & 4.1 &  4.3 &  4.5 &  4.7 \\
A      & 2.0  &    3.2 &  3.4 &  3.6 &  3.8 & (4.0) & 4.2 &  4.4 &  4.6 \\
       & 2.5  &    3.1 &  3.3 &  3.5 &  3.7 &  3.9 &  4.1 &  4.3 &  4.5 \\
       & 3.0  &   (3.0) & 3.2 &  3.4 &  3.6 &  3.8 & (4.0) & 4.2 &  4.4 \\
F      & 3.5  &    2.9 &  3.1 &  3.3 &  3.5 &  3.7 &  3.9 &  4.1 &  4.3 \\
       & 4.0  &    2.8 & (3.0) & 3.2 &  3.4 &  3.6 &  3.8 & (4.0) & 4.2 \\
       & 4.5  &    2.7 &  2.9 &  3.1 &  3.3 &  3.5 &  3.7 &  3.9 &  4.1 \\
G      & 5.0  &    2.6 &  2.8 & (3.0) & 3.2 &  3.4 &  3.6 &  3.8 & (4.0) \\
       & 5.5  &    2.5 &  2.7 &  2.9  & 3.1 &  3.3 &  3.5 &  3.7 &  3.9 \\
       & 6.0  &    2.4 &  2.6 &  2.8 & (3.0)&  3.2 &  3.4 &  3.6 &  3.8 \\
       & 6.5  &    2.3 &  2.5 &  2.7 &  2.9 &  3.1 &  3.3 &  3.5 &  3.7 \\
K      & 7.0  &    2.2 &  2.4 &  2.6 &  2.8 & (3.0) & 3.2 &  3.4 &  3.6 \\
       & 7.5  &    2.1 &  2.3 &  2.5 &  2.7 &  2.9 &  3.1 &  3.3 &  3.5
 \\
\hline
\end{tabular}
\end{center}
\end{table*}

\subsection{The observing programme}

We seek to target old stars in the thick disk and halo. In order to
target a large enough volume of the thick disk and halo, we need reach
to an apparent mag V = 17 (equivalent to I $\sim$ 16.5); see Table
\ref{tab2}.  We propose to observe stars in a narrow magnitude range,
15 $< V <$ 17, in order to reduce the effects of scattered light in the
instrument \citep{wg92,wg95}.

If we adopt Gilmore's model \citep{grh85} of the Galaxy
\citep[see also][]{rrdp03}, within our
magnitude range towards the Galactic poles, there are about 230 stars
deg$^{-2}$  with the following breakdown: 140 thin disk main sequence
(MS), 45 thick disk MS, 10 halo MS, 15 evolved (subgiant or red giant)
thick disk, 15 evolved halo, 5 halo horizontal branch. Along a cardinal
sight line of (l=90$^\circ$, b=30$^\circ$), these numbers increase to
820 stars deg$^{-2}$, with the following breakdown: 660 thin disk main
sequence (MS), 80 thick disk MS, 10 halo MS, 4 evolved thin disk, 43
evolved (subgiant or red giant) thick disk, 15 evolved halo, 5 halo
horizontal branch. This amounts to 60 thick disk stars towards the
poles, and 120 stars at the lower latitude; we observe about 30 halo
stars along either sight line.

A reasoned SNR calculation for a high-resolution KAOS design 
(see web calculator at http://www.noao.edu/kaos) shows that we should be able
to achieve SNR$\sim$100 per pixel at V = 17 in a 4 hr exposure
(R$\sim$20,000) rising to SNR$\sim$300 at V = 15. Here we assume a
0.5'' seeing observation of a 3000K blackbody source at 5000 ${\rm
\AA}$ in full moon. A median SNR$\sim$150 for the survey will be
sufficient to identify unique chemical tracks and chemical signatures.
This level of performance is encouraging for R$\sim$40,000 performance
if this could be factored into the KAOS design.

For sanity, we have checked the predicted performance against a UVES
commissioning observation (single slit, not fibre fed) which claims
SNR$\sim$100 at 4000 ${\rm \AA}$ for I$\sim$15 star at R$\sim$25,000 in
1 hour exposure. This is a factor of two better than KAOS as expected
 since KAOS is fibre fed.

The actual spectroscopic resolution we need to achieve depends in part
on the spectrophotometric stability, both in terms of an accurate zero
level to each spectrum, and an accurately calibrated instrument
response. If both zero level and instrument response can be calibrated
accurately and consistently, it may be possible to conduct the KAOS
survey at R = 20,000.  However, if, say, the zero level can be
maintained, but not the instrument response correction, it is safer to
work at R = 40,000, and fit individual lines in order to measure
equivalent widths. But the advantage of removing the instrument
response accurately is that it assists in determining the true
continuum level by matching to atmospheric models.  If more than one
order is observed, instrument response correction is also important for
joining orders together.

In summary, for the benchmark KAOS one degree field, we detect 
60,000 thick disk stars in 500-1000 KAOS fields, and 25,000 halo stars 
in about 1000 KAOS fields. Ideally, we would opt for 
R$\sim$40,000 but it may be possible to work at 
R$\sim$20,000 with sufficient instrument stability.

\section{Related surveys and the way ahead for near-field cosmology}

Detailed high resolution abundance studies of large samples of galactic
stars will be crucial to the success of the Galaxy Genesis project. The
KAOS survey is one of three massive stellar surveys.

The first is the RAVE survey (see http://www.iap.de/RAVE) which will
obtain radial velocities and abundances for 50 million stars, i.e. all
stars down to V=16, over the period 2005-2010 \citep{ms02}. This
requires a new generation 'Echidna' spectrograph mounted on a Schmidt
telescope (1.2m aperture), with 2250 fibres working at  a spectroscopic
resolution of  R $\sim$ 8000.

By far the most ambitious proposal is the GAIA satellite mission (see
http://www.rssd.esa.int/GAIA) over the period 2012-2018 which is
expected to launch during the time frame of the KAOS survey 
\citep{pdg01}. This will provide photometric, spectroscopic and phase
space information for a billion stars in the Local Group. The GAIA
mission is expected to obtain R $\sim$ 10,000 resolution spectroscopy
for up to 100 million stars or more.

It is important to note several points about the GAIA and RAVE surveys:
these are based on snapshot surveys with a 1m aperture telescope,
targeting the 8700\AA\ region at low resolving power. Their choice of R
= 8000 (RAVE) and R = 10,000 (GAIA) are supported by Figure
\ref{asspec2-fig}. As we have
seen, this region carries limited abundance information. Both missions
are primarily optimized for radial velocity studies. The KAOS echelle
survey will be the first million star survey optimized for chemical
abundance work.

The KAOS survey requires a highly multiplexed, high resolution
spectrograph which will be expensive and technically challenging, but
we believe this must be tackled if we are to ever unravel the formation
of the Galaxy.  Could some of the residual inhomogeneities from
prehistory have escaped the dissipative process at an early stage? We
may not know the answer to this question with absolute certainty for
many years.  But it is an intriguing thought that one day we may be
able to identify the Solar Family, i.e.  the hundreds or even thousands
of stars throughout the Gaiasphere that were born within the same cloud
as the Sun.

\end{document}